# Shifting 'AI Policy' Preprints and Citation Trends in the U.S., U.K and E.U., and South Korea (2015-2024)

**Simon Suh and Daniene Byrne, Ph.D.**

SUNY Stony Brook University
College of Engineering and Applied Sciences
Department of Technology and Society

The publication and subsequent citation of preprints are changing the distribution of scientific research globally. This study examines the citations of preprints on Artificial Intelligence (AI) policy, specifically focusing on the impact across two major disruptive events – the COVID-19 pandemic and the release of ChatGPT. It provides an annual comparison of patterns in the United States (U.S.), the EU (including the UK in Europe), and South Korea (Korea) from 2015 to 2024. Using bibliometric data from the Web of Science, global disruptive events are correlated with the adoption of preprints in AI policy research to examine whether or not outcomes vary by region. Analysis reveals that over the decade, all regions experienced significant growth in preprint citations, from below 5% to around 40% in 2024. The growth, magnitude, and trajectory of change varied slightly by region. This research provides evidence of a decade-long normalization of pre-prints across all three geographic and cultural regions, as well as in areas adjacent to AI, such as law and social science. A trend that follows in the footsteps of Computer Science, a field where preprint use is well established. Our outcomes emphasize the need for future AI Policy work, informed by researchers who understand the breadth of literature and can critically interpret quality preprints in the field of AI.

**Simon Suh, BS,** is an international scholar interested in comparative scientific responses to global events.

**Dr. Daniene Byrne** is a SUNY PRODiG+ Fellow at Stony Brook University. She studies policy challenges for emergent technologies and responsible AI policy. Orcid: 000-0001-5136-6983

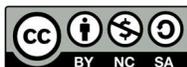

In the twenty-first century, the field of artificial intelligence (AI) experienced a revolutionary transformation. New applications and research are emerging at an incredible pace. This transition has not only accelerated the development of AI technologies but also reshaped the way scientific information is communicated and authenticated. We investigate two effects of this acceleration: the increasing demand for AI policy research and the tendency for scholars to publish and cite preprint publications. The term AI Policy encompasses both an ethical and legal framework applied in situations like academic use and in broader applications to AI governance protocols, which (ideally) are used to establish responsible AI protection and innovation standards. It is therefore deeply linked to AI development and an increasing global need for up-to-the-minute, shared research on AI understanding and regulation. Therefore, our investigation of the growing reliance on the citation of preprints focuses on articles with the keyword "AI policy".

Preprints, like other research products, share scholarly research theories, methods, and outcomes. But unlike most scientific journal articles, they are available to the public without undergoing a peer review process[1], [2]. Traditionally, highly esteemed peer-reviewed journals set the disciplinary standards for academic credibility (methods/theories, and reporting of findings) while acting as a filter to prevent low-quality research and ensuring the research's validity, significance, and originality[3]. They still play this prestigious role today. Unfortunately, this level of scrutiny requires a long publication process, which is not feasible in all fields or necessary for all types of research (for example, preprints successfully share methods related to computer code that can be run to determine their viability).

Certain research disciplines and academic communities accept the use of preprint platforms such as arXiv. These platforms allow them to share their research immediately. The growing shift in citations from reliance solely on traditional peer-reviewed publications to more frequent citing of non-peer-reviewed work posted on preprint platforms has raised critical questions about the continued reliability of academic research grounded in unreviewed research and the extent of preprint engagement. Preprints have been described as game-changing and paradigm-shifting[4], [5]and shaking up the traditional industry in academic publishing, potentially at the cost of harming scientific credibility. "Academics who cite invalid, poorly vetted, or false facts could cause harm, not unlike the unscholarly 'predatory' open access movement [5]Today, peer-reviewed journals are crucial in maintaining research quality and scholarly integrity. At the same time, in fields with rapidly emergent information and subject to change, preprints can be a necessity to ensure new findings can be shared and tracked[4], [6].

We explore the underlying citation of preprints in traditionally published research on the topic of "AI Policy" between 2015 and 2024, across three distinct global regions. We also examine how global disruptive events, the COVID-19 pandemic, and the release of ChatGPT correlate annually with the number of publications on AI Policy and the percentage of their arXiv preprint citation rates. Both events created significant global disruption, potentially leading to an environment where bad actors, inexperienced student researchers, and other untrained readers become aware of, access, and refer to preprints over more theoretically and methodologically grounded peer-reviewed publications. This shift could potentially increase the submissions of lower-quality research to these platforms. We explore this data across three major AI research regions: the United States (U.S.), Europe, and South Korea. These regions were selected for their differing cultures in governance of AI and innovation.

We hypothesize that there have been varying perspectives on the acceptance and use of preprints within AI Policy research, and that the response to preprint citations varies with global disruptions. We generated a multi-region preprint adoption curve to detect distinct regional differences and discuss what these differences mean for the normalization of preprint use in AI policy.

**Background**

Historically, preprints date back to the 1960s; they gained popularity in the 1990s with the introduction of an open-access archive known as arXiv[7]. Hosted by Cornell University, NY, arXiv hosts over two million open-access papers in fields such as physics, mathematics, computer science, biology, finance, and engineering[8]. In technical fields, preprints are far more accepted than in others. Computer science has always had a strong open-source movement where code is continually shared and updated. The preprint process for this field is a step toward formalizing this without the delay of more classic publishing. Preprints are citable and trackable, allowing scholars to stake a claim on their research while refining it for further publication. The Preprint process has the strength of bypassing the delays associated with peer review, as well as the weakness of bypassing academic and disciplinary standards (also associated with peer review). At its worst, preprints can lead to lapses in quality, including opportunities for unethical conduct such as falsification of authorship. If not regularly checked and corrected, this can undermine the integrity of the entire preprints system. Therefore, the value and role of pre-prints in advancing quality science are mixed.

As with Large Language Models, the successful use and relevance of preprints are largely dependent on the education and background knowledge of the scholar using them. For authors who use them, these tools help their research gain recognition and contribute to emerging dialogue as a more refined paper is developed. Following the success of arXiv, domain-specific platforms like bioRxiv and medRxiv emerged for life sciences and medicine. Prior to the COVID-19 pandemic, preprints were primarily used in fields like physics, computer science, and artificial intelligence (AI). Those in medical and biological fields were hesitant to use or cite these works due to concerns about credibility and misinformation[9], [10]. The COVID-19 pandemic disrupted this norm. Suddenly, health policymakers and researchers required rapid access to newly emergent data regarding the virus spread, symptoms, treatments, vaccines, and the efficacy of public health strategies.

**Factors Driving Preprint Adoption**

One factor behind this shift to preprints is the time involved in the peer-review process, which can take up to six months to two years due to multiple rounds of review and editorial stages[11]In a fast-evolving field like AI, this delay can significantly reduce the relevance of the research findings by the time they are formally published. In scientific areas where knowledge is updated daily, outdated publications have limited usefulness. Preprint platforms allow the timely sharing of findings by passing the formal peer review process. Accessibility and faster circulation are relevant factors for AI policy researchers, where emerging technologies can quickly outgrow the current regulatory framework.

Industry research and publication, for example, AI industries including Google, OpenAI, and DeepMind have released landmark models like GPT-3, DALL·E, and AlphaFold on preprint platforms. Leaving AI policy researchers, no choice but to cite important preprint documents. Arriving with industry legitimacy, their priority was speed and visibility.  Industry preprints might be considered academic advertising that not only release surprising innovations, but do so on their own terms, shaping public narratives. Preprints demonstrate a company's leadership in the AI sector, attracting top talent. Industry-led preprints have received substantial media coverage and citations even before formal review, suggesting their enormous influence[12]. Preprints also play a significant role in industry product development pipelines, enabling open feedback, downstream collaboration, and rapid iteration. Industry research outcomes can be implemented in real-world applications before they appear in peer-reviewed journals; a similar situation occurs in academic institutional commons that allow uploads of scholarly research published elsewhere[13].

The Open Access movement promotes access to high-quality, peer-reviewed work[15]. Not to be confused with the term "Open Science," which more broadly supports opening up access to research that may not always be vetted. The Open Access promotes transparency, inclusivity, and accessibility in scientific communication[14]  Open Access publishing (of peer-reviewed work) requires initiative from traditional publishing systems to remove barriers such as paywalls, aiming to make research freely available to the global public[15]. The increasing use of preprints reflects the "Open" in Open Science, while potentially misrepresenting the "Science".

Preprint platforms may appear to promote Open Science through their immediate access to research, encouraging early collaboration, and speeding up knowledge transfer to policy and practice, but their quality depends on authors' self-monitoring[10]. They provide a different, faster form of Open Science, which largely benefits experts by enabling them to distinguish the high-quality sources. Their  numerous shifting impacts on the greater scientific community and on interdisciplinary scholarship have been subject to study, but are not fully understood[15]–[18]. The rapid publication and immediate turnover time (with no barriers to entry or quality) have been questioned as a significant tradeoff due to the potential damage to the integrity of scientific research[4], [5]. As with AI usage, those exploring the ocean of preprint platforms such as arXiv, bioRxiv, and medRxiv benefit most when they already understand the broader context and methods (scholarship) of research in their field. The increasing availability of preprints is changing how peer-reviewed journal publishers operate and has led to more open availability of quality peer-reviewed Open Access articles, though often with accompanying publication fees[10].

This trend for rapid dissemination aligns with the positive benefits of the open research practices that emphasize transparency, collaboration, and knowledge sharing come with the increased risk of promoting lower quality work. There is a tradeoff between the speed and availability of a publication and the trustworthiness of the research within, that must always be critically examined on a paper-by-paper basis when evaluating the research and conclusions within preprints.  Glorifying preprints for their role in Open Science is unrealistic without also addressing the dangers of sloppy or, at worst, false scholarship to the scientific community.

The AI policy researchers find themselves in a position where they are both dependent on some preprints to understand the fast-evolving nature of AI. At the same time, grounded in social science methods of economics, public policy, history, and sociology, professions have a much slower academic turnaround and rely on data that sometimes takes months to generate. AI Policy makers are compelled to address preprints with skepticism and caution. Policymakers, regulators, and ethical boards must make informed decisions to ensure that governance frameworks are grounded in substantive and truthful research. Proponents of preprints may argue that the delay of peer-reviewed publications in such a fast-paced context can potentially lead to delays in implementing critical policy interventions, potentially allowing harmful technologies to be distributed to the public. The U.S. state and federal policy process largely lags far behind technological development, partly because regulators need time to understand the critical impacts of technological rollouts. Sometimes the laboratories of States provide room for comparisons. Another reason is the desire to have a light touch on regulation, promoting the unfettered growth of AI as something that is ultimately good for the nation and favoring industry interests in emergent AI policy[19]. At the same time, suggestions for internal policies related to emergent AI may be bound to rely upon preprints. Because the nature of federal policy making often invites public, including industry suggestions, through the notice and comment process, the information within preprints may be integrated into industry comments. Ideally, preprints keep regulatory bodies, researchers, and stakeholders informed about cutting-edge technical advancements and ethical debates as they unfold, with the important caveat that these groups can critically assess the quality of research.

Taken together, these factors —the time necessary for peer review, the appeal of open access, the strategic dissemination practices of AI industries, and the nature of policy making and demands to address emergent technologies —are all part of the complex reality of preprints. They highlight that, despite their limitations, preprints are a regular aspect of scientific publishing. Therefore, we hope to evaluate the extent to which they have become a regularly cited component of AI policy research.

**Citation Trends and Regional Differences in AI Policy Research**

The shifts in norms for speed and quality of publications vary across disciplines, and the acceptance of preprints, as with trust in AI and LLMs, is largely cultural. Therefore, we expect preprint adoption will vary regionally. In the broad research area of AI Policy, we examine three regional leaders in AI development, each with a different approach, to see if this is reflected in their preprint citation trends. he United States, Europe, and South Korea.

In the United States, the most recent Executive Order on Maintaining American Leadership in Artificial Intelligence encouraged collaboration between academia, industry, and government, with a focus on open-access dissemination and research speed and visibility[20]. Federal policy further reflects this shift, as documents like the Blueprint for an AI Bill of Rights[21], and the AI Risk Management Framework[22], [23], cite preprints and open-source materials. These references signal institutional trust and underscore the value of preprints for timely policymaking.

Europe favors peer-reviewed publications as the standard for research credibility and policy development. Yet, COVID-19 crisis pushed many European institutions to adopt preprints out of necessity. They became a vital tool for real-time communication. The European Open Science Cloud (EOSC)[24], launched in 2020, accelerated this shift by promoting open-access publication under FAIR principles—Findable, Accessible, Interoperable, and Reusable[18]. EOSC encouraged greater transparency and legitimacy for preprints across EU member states. Their major policy, the European Artificial Intelligence Act, has no preprint citations.

South Korea's National Strategy for Artificial Intelligence highlighted not only technological leadership but also the democratization of knowledge through open-access publishing[25]. AI Ethics Guidelines published by the Ministry of Science and ICT and the Korea Information Society Development Institute—explicitly cite preprints, signaling high institutional trust[26]. South Korea sees preprints as a mechanism for enhancing agility, innovation, and responsiveness in AI governance. Each country balances scientific authority, speed, and access. While the U.S. embraces rapid dissemination led by industry and open science, Europe appears to cautiously adapt within existing academic norms, and South Korea actively incorporates preprints into national Policy strategies.

**Risks**

Increasing use of preprints in scientific discussion and policy development also raises several concerns. They are not subjected to editorial standards, increasing the risk of methodological flaws, unsupported conclusions, or even misinformation[5]. They lack formal peer review, which compromises their credibility and reliability, and during COVID-19 led to glutted servers[17].  This concern is especially important in high-stakes fields like medicine, public health, and AI governance, where inaccurate information from preprints can cause a significant problem. During the COVID-19 pandemic, preprints were widely cited in the media and political discourse before undergoing any validation, contributing to a chaos of unverified claims at a time when public trust in science was decreasing. They are subject to misuse by non-expert audiences who cannot distinguish between high- and low-quality research, risking excessive credibility for unsubstantiated or unconfirmed results. The integration of preprints in AI policy aims to address emergent technologies, but policymakers should carefully manage it to ensure scientific credibility is not undermined. Unfortunately, this is more easily said than done.

We suspect that some of the same trends that impacted the field of medicine and computational sciences hold true for areas of governance, including the policy realm.  We expect the growing use of preprints as citations in AI policy research due to overlapping areas of concern for both Computer Scientists and AI policy scholars. When documents for describing and understanding computational methods are only in preprint form, then policymakers are forced to use and cite the work. In some cases, Computer Scientists, already familiar with preprints and even potentially preprint authors, are co-authors of important policy recommendations.

The increase in citation of preprints in papers on AI Policy may also indicate a shift in the academic acceptance of preprints as legitimate forms of scholarly communication rather than a passing trend.  An increase in industry production of papers related to AI Policy is expected as AI moves into every area of life, and the use of a neutral, public outlet such as arXiv for publication shares corporate research openly, putting it in the hands of scientists, policy makers, and the inquiring public[18].

## Methods

### Research Questions

We address two research questions: (1) To what extent are global disruptions, such as the COVID-19 pandemic and ChatGPT, correlated with the shift from peer-reviewed journals to preprint platforms in the area of AI policy research? (2) To what extent do these publication trends in preprint citation related to AI Policy vary across AI-leading regions such as the US, Europe, and South Korea?

### Research Design

This study adopts a cross-regional, annual analysis to evaluate how preprint citation trends in AI-policy research have shifted over time in response to two major global events: the COVID-19 pandemic and the release of ChatGPT. We focus on comparing patterns across the United States, Europe, and South Korea.

### Data Sources

We gathered bibliometric data from the Web of Science database (accessed through Stony Brook University), using the keyword "AI Policy" and applying a location filter for the United States, Europe, and South Korea (See Figure 1). Analysis focused on the number of AI policy-related papers that cited one or more arXiv preprints between 2015 to 2024, offering a consistent indicator of preprint adoption in policy-focused AI research. In this study, when we refer to Europe, we include publication data from a comprehensive list of countries classified geographically and politically as part of Europe or the UK (See Appendix A for the full list).

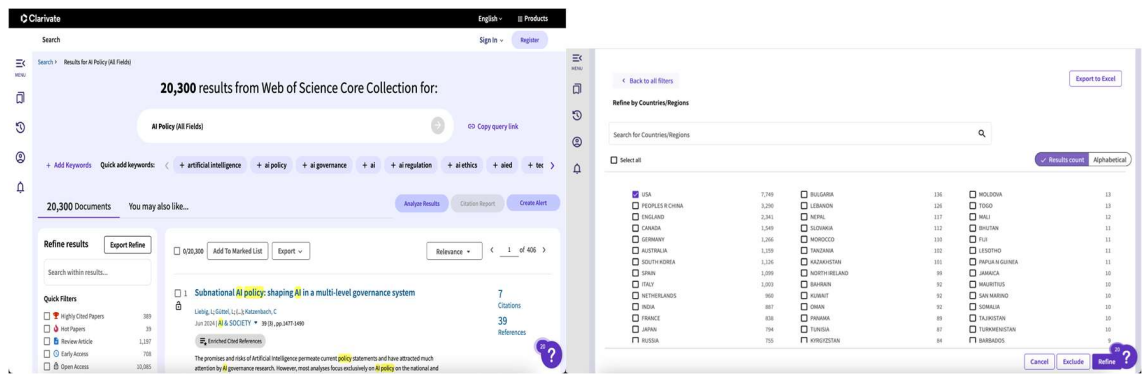

Figure 1: Keyword search of "AI Policy" in Web of Science (Left) / Location filter of United States with the same keyword in Web of Science (Right)

### Data Collection

Data was downloaded from the Web of Science interface for research published from 2015 to 2024. Data included full citation records for each publication with cited references, which were exported in BibTeX format. Due to Web of Science's export limit of 500 records per session, the exporting process was repeated until all the data files were exported. The final dataset resulted in:
- 6,252 records for the United States
- 6,165 records for Europe
- 1,045 records for South Korea

Then this data was filtered once again to include publications with an "arXiv" entry in their references, using R version 4.4.3.

**Data Analysis**

A cross-regional comparison evaluated changes in preprint citation trends on an annual basis for ten years from 2015 to 2024. This captured regional variations in response over time as compared to two key intervention points:

1. The onset of the COVID-19 pandemic (early 2020 )
2. The release of ChatGPT (late 2022)

These events were selected based on an expected impact on the publication numbers (we hypothesized a lull, then a surge due to the impact of COVID-19 on research and publication review. We also hypothesized another surge in articles with the release of ChatGPT and subsequent LLMs. Our analysis focused on identifying the total number of publications with the keywords "AI policy" that cited arXiv preprints before and after each event across the United States, Europe, and South Korea.

Within this dataset, the study identified papers citing (one or more) arXiv preprints, calculating the annual preprint citation rate as a percentage of total publications per region. This comparative approach allowed the study to track the shifts in citation behavior and assess regional differences in preprint adoption. By mapping these patterns around the two intervention points, the analysis explores how different institutional cultures and policy infrastructures shape the role of preprints in AI policy development and open science implementation.

**Limitations**

Our timeframe and access to publication databases limit this study. The use of a single database limited our understanding of region-specific publication trends, particularly for South Korea, which had the lowest number of records on Web of Science. In addition, we were limited by our inability to make month-to-month comparisons.

A comparative analysis using multiple scholarly databases—such as Google Scholar, ResearchGate, and the Directory of Open Access Journals (DOAJ) would have enhanced the scope of citation data and given an even clearer picture of the extent to which pre-publications are cited. This was not possible due to constraints such as paywalls and daily export limits. We attempted to gather publication data from South Korea's scholarly platforms, such as RISS, KCI, and DBPIA, and from European platforms, such as Open AIRE, CORE, and HAL.

Unfortunately, we encountered access restrictions and additional language barriers that complicated the search and retrieval processes.

**Future Research**

Future analysis will focus on AI Policy documents in the three regions, their citation of preprints, and specifically, which preprints are most cited by authors, as well as their rates of conversion. To better understand the role of preprints in governance. Research on the viability of creating a standardized reporting measure for peer-reviewed publications to list a preprint ratio might serve as an additional bibliometric indicator across several archive systems. An additional area of research is in the scholarly publishing community, focusing on studying the various responses to this trend and the trends of AI use in writing. Additional research on the literature covering the shifting opinions and scholarship on preprints over time will further inform this project. We also imagine further work in evaluating the quality of the highest performing preprints and the time scale before they are refined and accepted at more prominent journals.

**Results**

**Regional Trends Over Time (2015 - 2024)**

To understand how preprint usage has evolved across global AI research communities, this study analyzed bibliometric data collected from the Web of Science database, which includes "AI Policy" and cites at least one arXiv preprint. This analysis does not examine preprints that are hosted on arXiv directly, nor does it focus on official policy documents or government publications. Rather, the goal is to simply track citation behavior *within formal academic publications* and explore how different regions integrate arXiv preprints into scholarly communication on AI-related policy topics.

In addition to tracking citations of arXiv preprints, this study also analyzed the global peer-reviewed publication trends based on papers indexed in Web of Science for "AI Policy" (See Figure 2). This baseline enables a more accurate interpretation of whether the observed changes in preprint citation behavior are part of a broader increase in publication volume or represent a distinct regional shift toward open science dissemination. The time frame of 2015 to 2024 was selected to capture trends before and after two major global events: the COVID-19 pandemic (2020) and the release of ChatGPT (late 2022). By examining the annual change in the number of arXiv citations within this time frame, we identify the regional difference in how preprints have been adopted across the United States, Europe, and South Korea.

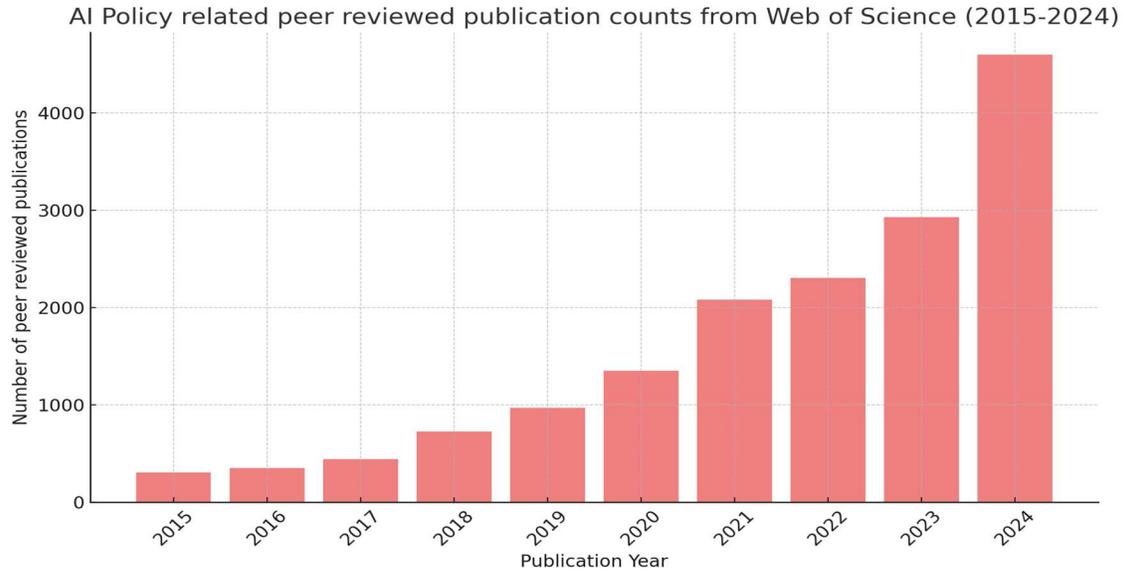

Figure 2: Total number of Peer-Reviewed Publications with the keywords "AI Policy" in Web of Science (2015-2024)

**United States**

There were 6,252 Web of Science-indexed publications in the United States containing the keywords "AI Policy" from 2015 to 2024. Out of those, 2,363 publications cited at least one arXiv preprint and were used to analyze the trend.

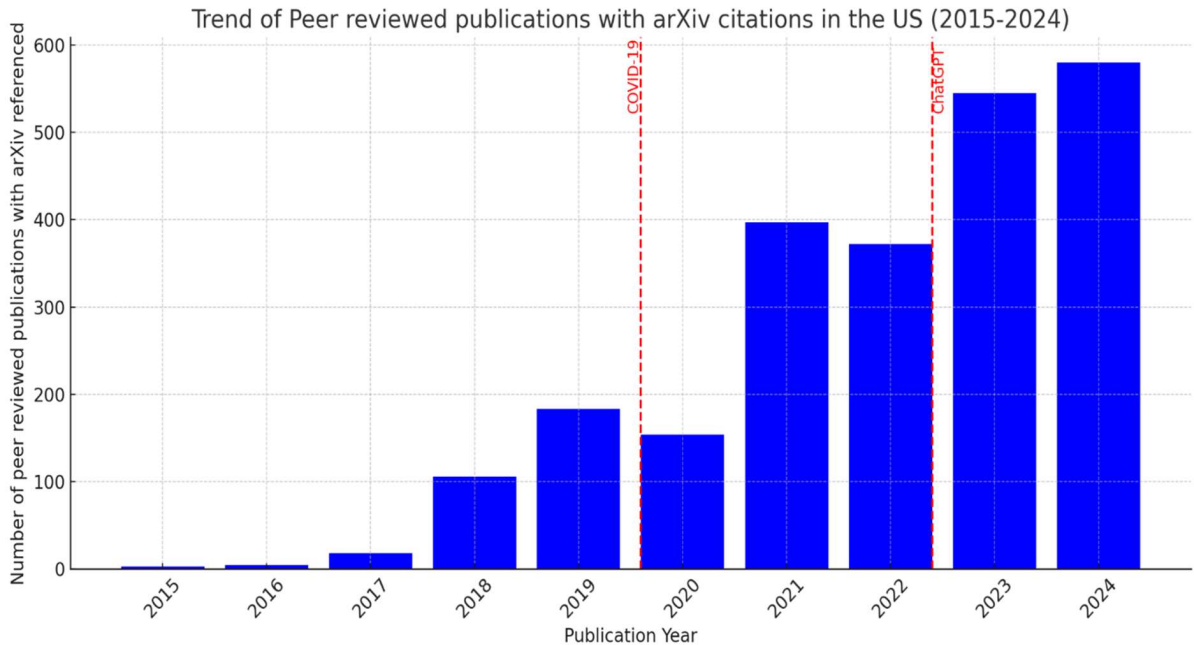

Figure 3: Number of arXiv-Referenced Peer-Reviewed Papers Published in the United States (Web of Science Analysis)

As shown in Figure 3, arXiv preprints citations rose modestly from 2018 to 2019 but then dipped slightly in 2020 – the year the COVID-19 pandemic began to disrupt research publishing globally. This short-term decline may reflect initial uncertainty and delays within academic publishing as institutions adapted to work remotely, and journal operations slowed.

After this brief decline, Table 1 reveals two big jumps in U.S. preprint citations. In 2021, the rate increased to 43.58%, marking the most significant single-year jump in the series, and peaking at 50.37% in 2023. This suggests that researchers increasingly turned to preprints as a faster and more accessible means of disseminating information with the release of ChatGPT and the COVID-19 pandemic's impact on publication workflow.

| Publication Year | Count of Publications with arXiv References in the US | Total Number of Peer Review Publications in US topic "AI Policy" | Annual arXiv citation rate in Peer Reviewed (%) |
|---|---|---|---|
| 2015 | 3 | 108 | 2.78% |
| 2016 | 5 | 184 | 2.72% |
| 2017 | 18 | 223 | 8.07% |
| 2018 | 106 | 338 | 31.36% |
| 2019 | 183 | 503 | 36.38% |
| 2020 | 154 | 541 | 28.47% |
| 2021 | 397 | 911 | 43.58% |
| 2022 | 372 | 827 | 44.98% |
| 2023 | 545 | 1082 | 50.37% |
| 2024 | 580 | 1535 | 37.78% |

Table 1: Annual Counts and Percentages of U.S. AI-Policy Publications Citing arXiv Preprints (2015-2024)

The delayed boost in U.S. citations indicates a broader shift, where preprints gained legitimacy. This pattern demonstrates a growing acceptance of arXiv within AI-related policy research, particularly as industry and academic researchers sought faster sources to influence public and regulatory discussions.

**European Regions**

Across the European Regions, there were a total of 6,615 publications containing the keywords "AI Policy" from 2015 to 2024, according to Web of Science. Out of those, 1,919 publications cited at least one arXiv preprint and were used to analyze the trend. As shown in Figure 4 and Table 2, preprint citation rates increased from 4.84% in 2015 to 6.11% in 2016, dipped slightly to 5.98% in 2017, and then accelerated to 15.83% in 2018. Rates further increased to 21.27% in 2019 and up to 21.96% in 2020 with the start of the pandemic. The most considerable boost occurred in 2021, when the rate jumped to 31.34%, reflecting the need for rapid dissemination during the pandemic. Following ChatGPT's release, Europe saw another significant increase to 40.88% in 2023. These two sharp increases demonstrate that European researchers not only responded to global disruptions but also promoted acceptance using preprint citations as standard practice. The increase in citations may also reflect delays in traditional journal publishing. As peer review took longer during the pandemic, preprints became an effective alternative for disseminating findings.

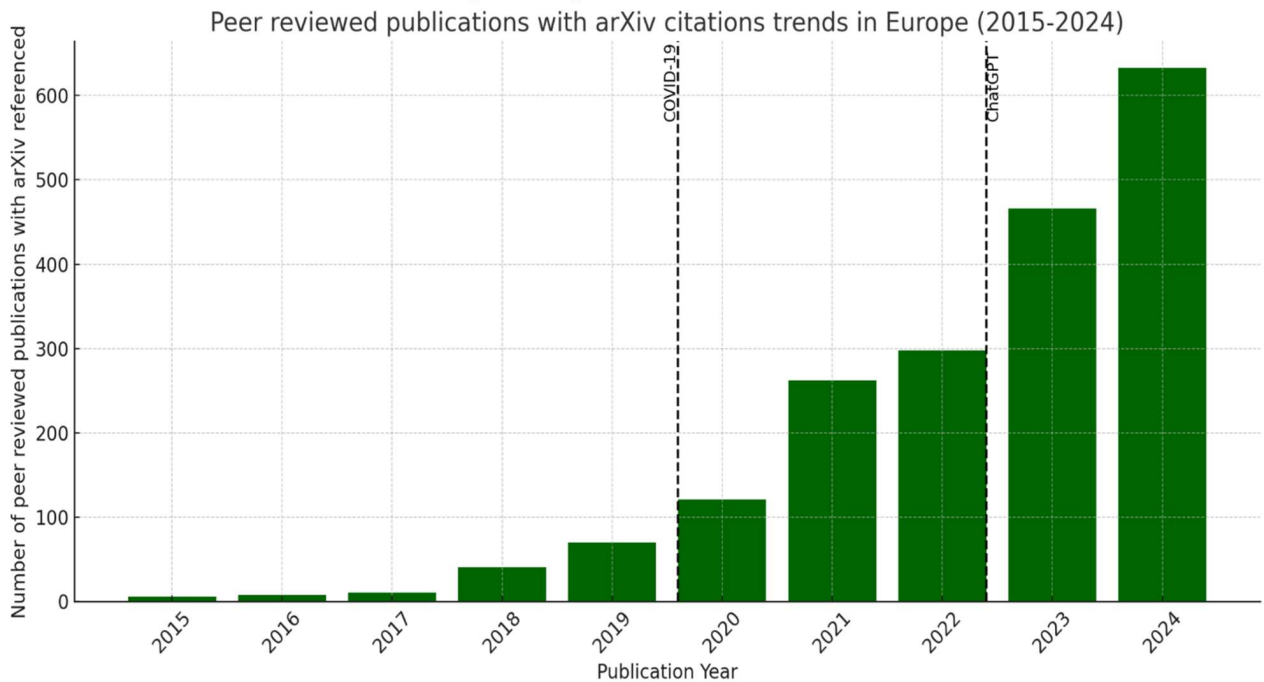

Figure 4: Number of arXiv-Referenced Papers Published in Europe (Web of Science Analysis)

Europe's growing use of preprints during this time also aligns with broader Open Science initiatives, which encouraged transparency and accessibility even during periods of disruption.

| Publication Year | Count of Publications with arXiv References in Europe | Total Number of Peer Review Publications in topic "AI Policy" | Annual arXiv citation rate in Peer Reviewed (%) |
|---|---|---|---|
| 2015 | 6 | 124 | 4.84% |
| 2016 | 8 | 131 | 6.11% |
| 2017 | 11 | 184 | 5.98% |
| 2018 | 41 | 259 | 15.83% |
| 2019 | 70 | 329 | 21.27% |
| 2020 | 121 | 551 | 21.96% |
| 2021 | 262 | 836 | 31.34% |
| 2022 | 298 | 878 | 33.94% |
| 2023 | 466 | 1140 | 40.88% |
| 2024 | 633 | 1647 | 38.43% |

Table 2: Annual Counts and Percentages of Europe AI-Policy Publications Citing arXiv Preprints (2015-2024)

**South Korea**

South Korea had a total of 1,123 publications containing the keywords "AI" AND "Policy" from 2015 to 2024, according to Web of Science. Out of those, 395 publications cited at least one arXiv preprint and were used to analyze the trend. As shown in Figure 5 and Table 3, citation rates were 0% through 2017, then climbed to 12.5% in 2018 and 20.5% in 2019. The pandemic period saw a further rise to 24.7% in 2020, but the most dramatic acceleration occurred in 2021, when the rate jumped to 41.3%. Unlike the U.S. and Europe, South Korea did not experience a post-ChatGPT spike in preprint citation trends. Instead, its preprint citation trend followed a consistent upward linear trend, indicating that open-access dissemination through arXiv may have already been integrated in the institution before the external shocks occurred. This stability supports the idea that South Korea has adopted preprints as part of a broader national strategy to promote open science, as highlighted in its National Strategy for Artificial Intelligence. Rather than reacting to external disruptions, the growth of preprint citations in South Korea appears to reflect a long-term institutional commitment to accessible and quick research dissemination.

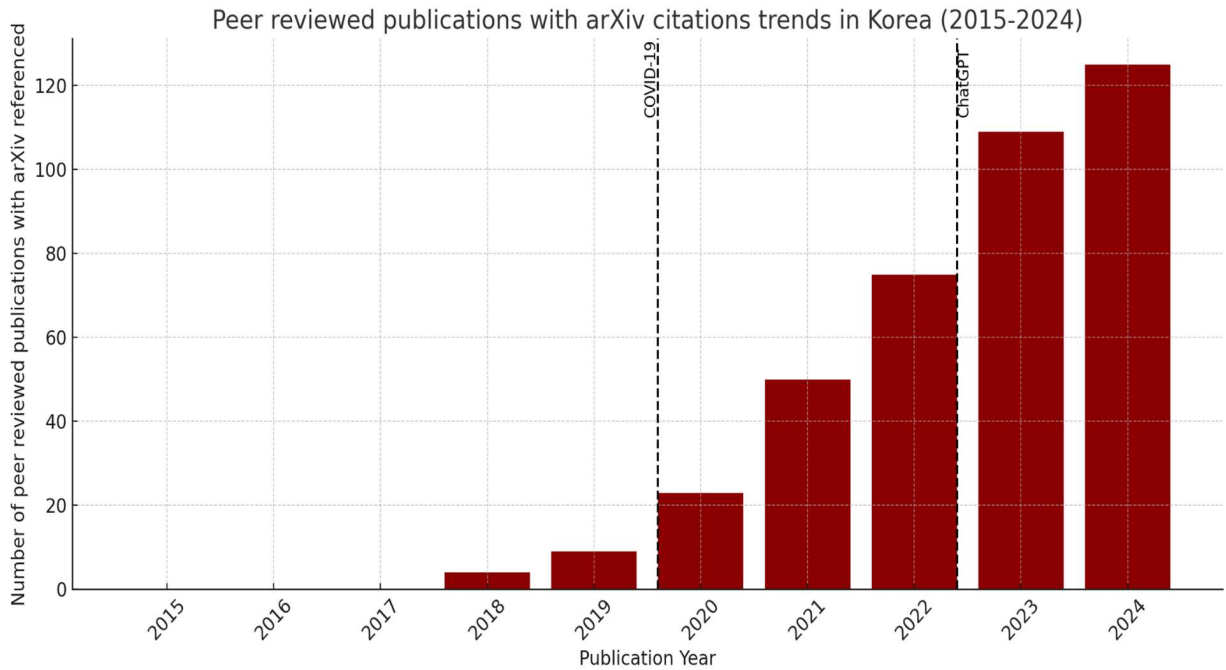

Figure 5: Number of arXiv-Referenced Papers Published in South Korea (Web of Science Analysis)

| Publication Year | Count of Publications with arXiv References in South Korea | Total Number of Peer Review Publications in South Korea, topic "AI Policy" | Annual arXiv citation rate in Peer Reviewed (%) |
|---|---|---|---|
| 2015 | 0 | 9 | 0% |
| 2016 | 0 | 15 | 0% |
| 2017 | 0 | 20 | 0% |
| 2018 | 4 | 32 | 12.5% |
| 2019 | 9 | 44 | 20.45% |
| 2020 | 23 | 93 | 24.73% |
| 2021 | 50 | 121 | 41.32% |
| 2022 | 75 | 172 | 43.6% |
| 2023 | 109 | 243 | 44.86% |
| 2024 | 125 | 296 | 42.23% |

Table 3: Annual Counts and Percentages of South Korea AI-Policy Publications Citing arXiv Preprints (2015-2024)

**Regional Trends in Citations of Preprints Over Time**

The decision to examine regional trends over time was made to evaluate whether the adoption of open science practices, such as citing arXiv preprints, was correlated with two major external interventions - the onset of the COVID-19 pandemic and the release of ChatGPT - or whether it reflected distinct regional dynamics. By comparing the United States, Europe, and South Korea, we examine the global influence of these disruptive events and the extent to which regional differences lead to the adoption of open science norms. Understanding these patterns increases our understanding of research publication practices across diverse international contexts.

The regional trend analysis (Figure 7) reveals differences in the arXiv citation rate for AI Policy publications across the United States, Europe, and South Korea. In the United States, it grew steadily from approximately 10.5% in 2018 to around 28.5% in 2020, reflecting the impact of the pandemic. It then significantly increased to 43.6% in 2021, and further to 50.4% by 2023, following the public release of ChatGPT. At this point, half of the US citations within papers on AI Policy were from arXiv preprints. This has since returned to around 40%.

In Europe, a similar but more gradual trajectory emerged. Citation rates increased from approximately 9.2% in 2018 to 21.4% in 2020, continued to rise to 31.3% by 2021, and reached 40.9% in 2023. After ChatGPT's release, the rate rose to 48.1% in 2024, indicating a consistent embrace of preprints. While the growth was less event-driven than in the U.S., the overall acceleration suggests that institutional efforts—such as the European Open Science Cloud (EOSC) and FAIR principles—played a significant role in shaping sustained adoption across the EU.

In South Korea, the citation rate grew in a more linear fashion. From 12.5% in 2018, it increased to 33.3% in 2020, rose to 39.4% in 2022, and peaked at 44.9% in 2023. Unlike the U.S. and Europe, South Korea's growth pattern was not characterized by sharp post-event surges. Instead, it reflected a longer-term strategy toward open science, as outlined in its 2019 National AI Strategy. Although South Korea's overall publication volume is smaller, the relative adoption rate of preprints is comparable to other regions.

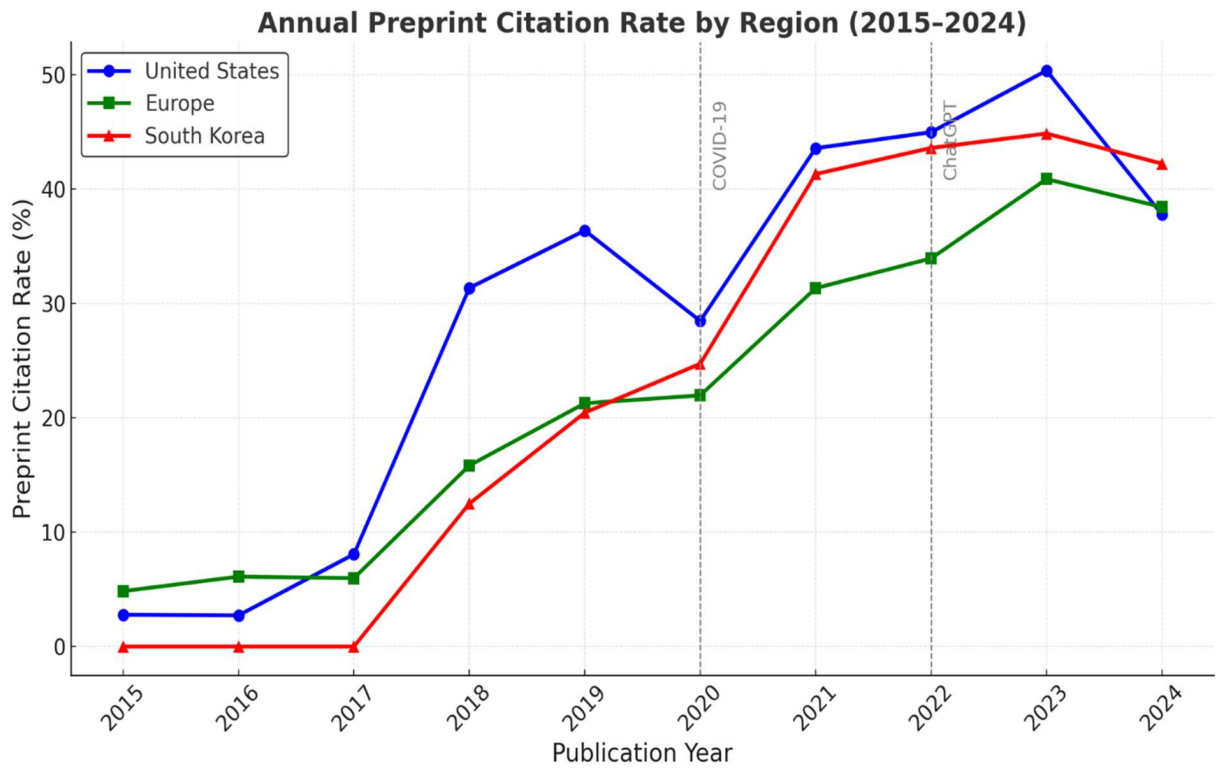

Figure 6: Comparative Annual arXiv Preprint Citation Rate for U.S., Europe, and South Korea

**Findings**

In answer to our research question, to what extent do these publication trends in preprint citation related to AI Policy vary across AI-leading regions such as the US, Europe, and South Korea? We found a publication trend across all regions, both in the number of publications including "AI Policy" and the number of citations of arXiv preprints in these publications, which have accelerated globally. The magnitude and pattern of change vary slightly across regions. Comparatively speaking, the U.S. percentage of preprints in all years was higher than those of Europe and Korea and surged earlier. However, as of 2024, they appear to be leveling out at a rate of forty percent of publications citing preprints.

In response to the question: To what extent are global disruptions, such as the COVID-19 pandemic and ChatGPT, correlated with the shift from peer-reviewed journals to preprint platforms in the area of AI policy research? We found that the U.S. publication rate and adaptation of preprints were much more sensitive to the shocks of COVID-19 and the release of ChatGPT. In the United States, in 2020, during the COVID-19 pandemic, both the number of publications and the percentage of arXiv preprints decreased relative to 2019. The same pattern occurred, but slightly more so, after the release of ChatGPT. On the other hand, Europe's publication and citation of preprint trajectories (within Web of Science citing from arXiv) were both steadily upward. Similarly, South Korea displayed an even more pronounced upward pattern of uninterrupted growth in preprint citations.

This may be an artifact of the dataset's source and the significantly larger number of data points in the US data, especially when compared to South Korea, a much smaller country with numerous other publication platforms. One explanation for the pattern in the United States between 2019 and 2020 is the slowing of review processes, which caused a backlog of papers and resulted in fewer preprint citations due to reduced overall publishing. The dip is followed by a bump of activity with 2021 moving above 2019 levels. Excluding the dip of COVID-19, and the smaller one of ChatGPT in US data, one can imagine a gradually increasing trajectory in the citation of arXiv preprints that begins in 2018 and increases at about 4-5% annually until reaching 50% in 2023. It is possible that ChatGPT's release, followed by a minor dip in 2022, might result from a slightly more critical analysis of preprints. This would require further study. The US preprint appears to reflect a saturation point at around 50% in 2023, with a slight decrease in 2024. We will have to wait for the 2025 data to be more certain.

Regional differences accurately reflect local research cultures or more clearly reflect only trends within Web of Science and arXiv. Both are centered in the US, so they may capture the most accurate picture of the US. In the case of Korea, it is difficult to determine if the trend we see in Web of Science publications is similar to what occurs in other publication databases open to the Korean language, and hosted in Korea. Is our result simply a trend of South Korean researchers publishing in US-indexed publications at increasing rates and citing US preprint sources.

Although it may be convenient to tie up the preprint trends to local research cultures, policy environments, and levels of open science maturity. Without further investigation of other preprint databases, we conclude we do not have sufficient evidence to make this claim.

**Implications**

We selected the topic of AI Policy, as governance of AI is a global issue emerging as AIs become more available and reliable tools. It not only spans countries, but it also spans disciplines, since so many are using AI for so many different applications, which require different kinds of policy solutions. The need for policy and the scholarly concern across numerous disciplines made it an ideal topic for this multinational comparison. A successful pre-print citation requires an understanding of the topic and domain knowledge to ensure the work is valid. Paired with the release of ChatGPT, as a language and an eventually paper-generating tool, a contemporary researcher must be doubly cautious of the scholarship behind pre-print forms of research.

In rapidly advancing fields, such as computer science, preprints and conference proceedings ensure researchers can keep pace with technological change. In the case of AI policy, which must keep up with changing technological advances and balance that with accurate and applicable outcomes, it is perhaps no coincidence that pre-print adoption levels surged to around 50% in 2023 and may level off. Preprints are now a large part of cited literature, and clearly, this is changing the way scholarly research is shared and interpreted across the globe. It is striking how, besides the more rapid adaptation of preprints by the U.S. in 2017-2018, there were not great regional differences in preprint adaptation. All usage has increased and leveled at around 40%. Demonstrating that researchers globally have normalized pre-print use (in the case of AI Policy research) on a global scale. This may both reflect changing attitudes toward peer review and a greater acceptance of pre-publications, as well as the need to rely on pre-prints as the best source on the topic. It is important that those interested in AI policy research, and anyone working in research reliant on preprints, read cited works closely and ensure their rigor. Failing to do so lowers standards and threatens research quality and trust in the scientific community.

**Limitations**

Our work did not capture the differences in regional cultures as much as we expected. We were limited to the Web of Science database, which may not capture the full spectrum of preprint adoption behaviors, particularly excluding non-English or regional publications. Our focus on citation behavior did not assess the content or quality of either the cited papers or preprints themselves.

Finally, in the interests of transparency, and in a nod to combining our form and content, aspects of our writing process, particularly organization of the paper, and literature initial suggestions by ChatGPT were supplemented and superseded by more rigorous critical work. We examined an element of bias favoring preprints as part of a noble quest toward Open Science and their portrayal as a future-looking panacea for the global advancement of transformational research and AI policy. We expect this misrepresented the values of Open Access, which are about opening up peer-reviewed work to the public[15].  A comprehensive study of terms, Open Access, Open Science, and their implications in the preprint debate were beyond our scope and would require further research. Though we do welcome comments and hope our study sheds light on one corner of the issue related to policymaking for AI.  Notably, we initially cited one preprint as an ArXiv paper, "Big Tech Influence over AI research revisited" 2023[12]. However, upon further investigation, it has since been published in the Journal of Infometrics. Several of the cited papers in ArXiv may also have been accepted for publication by the time policy research papers were published, though it is difficult to distinguish if this was known. We are also curious: would authors upload earlier versions of accepted papers as preprints, as a way to further disseminate their research?

**Suggestions for Future Research**

Our research demonstrates that the citation of preprints is becoming normalized in research cultures across the U.S., E.U., and South Korea.  Expanding this scope to include Canada, South America, Africa, the Middle East, Southeast Asia, India, and China would provide a more inclusive and comprehensive global picture. The increased production of research papers related to AI Policy, reflects a global call for research in AI Governance across a range of topics, from overseeing LLM applications and use to governing AI model architecture, data use, and collection to the overall mission and end-user experience of a given AI.

Since preprints do become published in normal journals, it would be interesting to know the conversion rates over time across these same three regional groups. We could also extend the data collection period beyond 2025 to better capture long-term post-ChatGPT trends and evaluate whether the observed peaks and stabilization in preprint citations are sustained or temporary. Another future topic of study is to look at all major policy documents across the three regions to evaluate their citation sources. One could also compare the spread across industry, government, and academia. This might be highly beneficial as a point of comparison in how work across these three groups is cited and potentially becomes influential.  Additionally, prompting Open Science and preprint issues to various reasoning LLMs might provide data for AI policy bias studies.

Informational interviews with journal publishers of peer-reviewed journals might provide an interesting perspective on the way peer review is working to address preprints and open access. We have noticed that some publishers regularly survey scholars to understand their research needs and perspectives, and that some preprint platforms like arXiv encourage authors to link preprint papers to their eventually peer-reviewed accepted counterparts.

Often, industry papers are posted directly to sites like arXiv or to their own webpages. In addition to expanding the temporal scope, a more in-depth analysis would be beneficial. For example, future research could incorporate multiple bibliometric databases such as Google Scholar, CORE, and national preprint repositories to enhance the reliability and generalizability of the results. We could also download, open, and sort individual preprints by month of publication while gathering data from these bibliometric databases. This would allow for a detailed breakdown of annual publication volumes into monthly ratios, offering a deeper insight into the dynamics of preprint adoption around key intervention points such as the COVID-19 pandemic and the release of ChatGPT. Finally, gathering quantitative data with qualitative research, such as interviews with researchers, policy advisors, and journal editors, could provide valuable context regarding the motivations, attitudes, and institutional barriers influencing preprint usage across different regions.

**Conclusion**

Mid 2025 is a time when AI Policy within the US, EU, and South Korea, as well as globally, is in flux and evolving rapidly. The question of which research informs policy and the quality of this research is critical. Our work on AI Policy preprint citations from 2015 to 2024 has found that preprint publications have moved from a fringe use to a regular source of citations. This supports findings that preprints have become a primary source for research[2]. Without surveying researchers, it is difficult to say if this is due to comfort in the process or more of a necessity. But preprints can provide a place for early support and feedback.

We observed a steady increase in preprint citations globally across all regions, with the U.S. having the highest percentage of preprint citations. While Europe and South Korea maintained a steady and consistent rise to around forty percent. It now appears AI policy preprint citations could level off, but it is difficult to tell until we have 2025 data. The U.S. experienced the most significant dip in both publications and the percentages of preprints in 2020. This was not reflected in trends in Europe or Korea, which each maintained more consistent, linear growth in preprint adoption. The increasing influence of AI, both before and after the release of ChatGPT, had no marked impact on the constant and upward trajectory of AI Policy publication output in Europe or Korea. However, it did correlate with a slight dip in publications related to AI Policy in the US. Robust data for the U.S. response to COVID-19 results suggest there may be other web resources (outside Web of Science) where a greater majority of European and UK, as well as Korean scholars, are more well represented, and therefore, it is difficult to determine if this is an actual trend or an effect of our use of only Web of Science data.

AI Policy researchers must rely on pre-prints on some topics as the basis for shaping or imagining AI policy. This research records a striking time when AI is increasingly used across all fields of research, writing, and professions, sparking a demand for quality AI policy and generating a surging number of publications in response to this need. The full drivers of this trend and outcomes are still not fully understood. In addition, preliminary work on this paper had ChatGPT suggest that our research hinges on the concepts of Open Science, a noble theory that glorifies preprints as a means of opening up technological and scientific research advancements to the world. This was suggested as a theory worth shaping this paper around, while dismissing research addressing the problem of preprint falsification, shoddy research, and biased perspectives. Promoting the uploading and citation of sources on preprint platforms, which are open not only to researchers but also to inclusion in AI datasets, amplifies errors and expands the scope of the LLMs that feed off them.

Researchers recognize that scientific research and writing require human time and effort, as they are not generated from one-off questions to reasoning AI, later uploaded as scholarship. The danger of having no boundaries includes a lack of standards and creating confusion for the uninformed or non-specialized reader. We dub the practice of uninformed, naive people who misunderstand scholarship, taking advantage of open access to submit anything for a citation, as "enfalsification." A play on Corey Doctorow's term of art, first used to refer to the realm of social media[27]. Peer review, like all critical human analysis, deserves recognition and compensation and requires grounding in disciplinary norms and awareness of the broader scientific context. But not everyone has the experience or benefit of an able mentor and reviewer. Not all preprints are created equally, and for the uninformed, this can be difficult to discern. Yet, they have their place, making space to float work in transition, sharing recent findings, and preparing for print versions, they provide a useful (if temporary) landing space.

   We are facing a new age where it is tempting to outsource research and analysis to AI. Our need for AI policy based on research that is both high quality and timely is now more critical than ever.  We hope this work both highlights the normalization of non-peer-reviewed work and provides citation data up to 2024 as a benchmark, and as grist for critical discussion of the potential harms of the current preprint usage trends. We cannot overstate the importance and practicality of recognizing that a good portion —still around sixty percent of citations —are sourced from peer-reviewed research, which grounds AI policy and science on what we believe to be a solid and reliable research foundation.

# Appendix A

**European regions included in Web of Science dataset**
**(including EU member states and other geographically associated countries)**

England, Germany, France, Spain, Italy, Netherlands, Switzerland, Sweden, Belgium, Norway, Finland, Portugal, Poland, Greece, Austria, Ireland, Romania, Denmark, Hungary, Estonia, Latvia, Lithuania, Bulgaria, Croatia, Slovenia, Slovakia, Ukraine, North Macedonia, Serbia, Albania, Moldova, Bosnia Herceg, Montenegro, Kosovo, Iceland, Malta, Cyprus, Luxembourg, Liechtenstein, Monaco, San Marino, Scotland, Wales, Northern Ireland

[20] *Intelligence*. 2019.

[21] The White House, "Blueprint for an AI Bill of Rights - MAKING AUTOMATED SYSTEMS WORK FOR THE AMERICAN PEOPLE," *White House*, no. October, pp. 1–73, 2022, [Online]. Available: https://www.whitehouse.gov/ostp/ai-bill-of-rights.

[22] R. Schwartz *et al.*, "The Draft NIST Assessing Risks and Impacts of AI (AIRA) Pilot Evaluation Plan," 2024.

[23] NIST, "Artificial Intelligence Risk Management Framework : Generative Artificial Intelligence Profile," *Natl. Inst. Stand. Technol.*, 2024, [Online]. Available: https://doi.org/10.6028/NIST.AI.600-1.

[24] European Commission, *Proposal for a REGULATION OF THE EUROPEAN PARLIAMENT AND OF THE COUNCIL LAYING DOWN HARMONISED RULES ON ARTIFICIAL INTELLIGENCE (ARTIFICIAL INTELLIGENCE ACT) AND AMENDING CERTAIN UNION LEGISLATIVE ACTS*. European Union, 2021.

[25] "National Strategy for Artificial Intelligence," Korea, 2019 *https://www.msit.go.kr/bbs/view.do?bbsSeqNo=46&mId=10&mPid=9&nttSeqNo=9&sCode=eng.*

[26] AI ETHICS COMMUNICATION CHANNEL, "National Guidelines for Artificial Intelligence Ethics, Korea, AI Ethics Framework Enhancement Project," *https://ai.kisdi.re.kr/eng/main/contents.do?menuNo=500011*, 2025. .

[27] C. Doctrow, *Enshittification*. Farrar, Straus & Giroux MCD x FSG Books, 2025.